\documentclass[FBSedit,FBSmath,ecsub]{FBSsuppl}
\usepackage{cite}
\usepackage{amsfonts}
\usepackage{amssymb}

\renewcommand{\deg}{^{\circ}}

\newcommand{\eqnb}{\begin{equation}}
\newcommand{\eqne}{\end{equation}}

\newcommand{\question}[1]{}
\newcommand{\NSt}{{\mbox{\scriptsize\it NS}}}
\newcommand{\St}{{\mbox{\scriptsize\it S}}}
\newcommand{\slashed}[1]{\rlap{$#1$}/}

\title{ Schwinger-Dyson approach to light pseudoscalars }
\author{D. Klabu\v{c}ar$^a$ and D. Kekez$^b$}
\institute{$^a$Physics Department, Faculty of Science, Zagreb University,
Bijeni\v cka c. 32, HR-10000 Zagreb, Croatia  \\
$^b$Rudjer Bo\v{s}kovi\'{c} Institute, P.O.B. 180, HR-10002 Zagreb, Croatia}

\begin{document}

\maketitle
\begin{abstract}
We review briefly some of the successes of Schwinger-Dyson (SD) approach 
to the physics of quarks and hadrons, primarily light pseudoscalar mesons
including $\eta$ and $\eta'$. 
The main purpose of this paper is to point out that SD results 
on $\eta$ and $\eta'$ mesons can be formulated and given also in the
two-mixing-angle scheme.
\end{abstract}

\noindent
Schwinger-Dyson (SD) equations provide a nonperturbative, covariant 
and chirally well-behaved approach to quark-hadron physics. This 
approach \cite{recentReviews,Alkofer:2000wg,Roberts:2000aa}
can be formulated so that it has strong and clear connections with QCD. 
Admittedly, coupled integral SD equations for Green's functions of QCD 
cannot be solved exactly since 
some truncation of the infinite system of SD equations is unavoidable.
However, the numerous advances 
\cite{recentReviews,Alkofer:2000wg,Roberts:2000aa}
achieved in this field within the last 
dozen or so years also include increased understanding of how to perform
this truncation so that some important known behaviors are preserved.
This includes the known perturbative QCD behavior 
and dynamical chiral symmetry breaking (D$\chi$SB).
The latter is crucial in low-energy QCD, especially for the description 
of the light-quark sector. The correct chiral behavior
due to D$\chi$SB can be achieved through the coupling of
SD and Bethe-Salpeter (BS) equations in a consistent approximation.
That is, one usually solves 
\begin{equation}
S^{-1}_q(p) = \slashed{p} - m_q
      - i g^2 \, C_F \int \frac{d^4k}{(2\pi)^4}
       \gamma^\mu S_q(k) \gamma^\nu G_{\mu\nu}(p-k)~,
\label{SDE-rainbow}
\end{equation}
\noindent 
the (rainbow-)ladder-approximated SD equation for ``dressed" propagators, 
$S_q(p)$, of the light quarks $q=u, d, s$. 
The strong coupling constant is denoted by $g$ and the color factor $C_F=4/3$.
The interaction $G_{\mu\nu}$ stands for an effective 
gluon propagator; it is known in the perturbative 
QCD regime, but has to be modeled in the nonperturbative regime, for low 
momenta, in order to be phenomenologically successful. The same ``dressed" 
gluon exchange interaction $G_{\mu\nu}$ and the resulting solutions of 
Eq. (\ref{SDE-rainbow}) for light-quark propagators $S_q$, are then employed in 
\begin{equation}
      S_q^{-1}(k+\frac{P}{2})
      \chi_{q{\bar q}'}(k,P)
      S_{q'}^{-1}(k-\frac{P}{2})
      =
      i g^2  C_F \int \frac{d^4k^\prime}{(2\pi)^4}
      \gamma^\mu
      \chi_{q{\bar q}'}(k^\prime,P)
      \gamma^\nu
      G_{\mu\nu}(k-k^\prime)~,
\label{BSE_MJ}
\end{equation}
the ladder-approximated BS equations for various quark-antiquark ($q\bar q'$) 
BS amplitudes $\chi_{q{\bar q}'}$. This defines a chirally well-behaved SD-BS 
approach,
in which light pseudoscalar meson $q\bar q$ bound states ($\pi^{0,\pm}, 
K^{0,\pm}, \eta$) in the chiral limit (and {\it close} to it) {\bf simultaneously}
manifest themselves also as ({\it quasi-})Goldstone bosons of D$\chi$SB.
This enables one to work with the mesons as explicit $q\bar q$ bound states 
while reproducing analytically -- in the chiral limit -- the famous 
results of the Abelian axial anomaly for the light pseudoscalar mesons, 
namely the amplitudes for $\pi^0\rightarrow\gamma\gamma$ and 
$\gamma \rightarrow \pi^+ \pi^0 \pi^-$ \cite{Alkofer:1995jx}.
This is unique among the bound state approaches -- {\it e.g.}, see
Refs. \cite{Alkofer:2000wg,Roberts:2000aa,Kekez:1998xr,Alkofer:1995jx} 
and references therein. Nevertheless, one keeps the advantage 
of bound state approaches that from the $q\bar q$ substructure
one can calculate many important quantities such as meson decay 
constants ({\it e.g.}, $f_\pi$), which are just 
parameters in most of other chiral approaches to the light-quark sector. 
Also, one can depart far from the chiral limit, soft point and on-shell limit 
in the anomaly-related processes $\pi^0,\eta,\eta' \leftrightarrow\gamma\gamma$ 
and $\gamma \rightarrow \pi^+ \pi^0 \pi^-$, for example in Refs.
\cite{Kekez:1996az,Klabucar:1997zi,Kekez:1998xr,Kekez:1998rw,Kekez:2000aw,Klabucar:2000me,Kekez:2001ph,Bistrovic:1999dy} 
where the phenomenologically successful Munczek-Jain Ansatz \cite{Jain:qh} 
for the effective gluon propagator $G_{\mu\nu}$ was used.
(This Ansatz is very similar to the interactions adopted, with small 
variations, in almost all recent phenomenologically most successful SD 
studies \cite{recentReviews,Alkofer:2000wg,Roberts:2000aa}.)
In this way obtained description of $\eta$--$\eta'$ complex 
\cite{Klabucar:1997zi,Kekez:2000aw,Klabucar:2000me,Klabucar:2001gr,Kekez:2001ph}
is especially noteworthy, as it is very successful in spite of the limitations 
of the SD-BS approach in the ladder approximation, because of which Eqs. 
(\ref{SDE-rainbow})-(\ref{BSE_MJ}) cannot include the effects of gluon anomaly. 
They give only non-anomalous contributions to $q\bar q$ BS solutions and 
masses. Fortunately, gluon anomaly is suppressed as $1/N_c$ in the expansion
in number of colors $N_c$, so that the procedure of Refs. 
\cite{Klabucar:1997zi,Kekez:2000aw,Klabucar:2000me, BledWorkshop2002}, which includes
gluon anomaly only through parameterizing the anomalous shift of $\eta_0$ mass,
leads to very satisfactory results. For example, our $\eta$--$\eta^\prime$ 
mass matrix is in agreement with phenomenology and lattice results.
It can be seen it leads to a very good description of 
$\eta$--$\eta^\prime$ mixing, although it is formulated 
(except in Appendix of Ref. \cite{Kekez:2000aw}) in 
terms of one {\it state} mixing angle,
for example $\phi$, 
\begin{equation}
|\eta\rangle = \cos\phi |\eta_\NSt\rangle
             - \sin\phi |\eta_\St\rangle~,
\,\,\,\,\,\,\,
|\eta^\prime\rangle = \sin\phi |\eta_\NSt\rangle
             + \cos\phi |\eta_\St\rangle~,
\label{eqno3}
\end{equation}
rotating the nonstrange-strange ({\it NS-S}) basis states $\eta_\NSt$ and $\eta_\St$ 
to the mass eigenstates $\eta$ and $\eta'$. In terms of the quark-antiquark
states $|q\bar{q}\rangle$ ($q=u,d,s$), or the (effective, flavor-symmetry-broken) 
singlet-octet states $\eta_8$ and $\eta_0$,
        \begin{eqnarray}
        |\eta_\NSt\rangle
        &=&
        \frac{1}{\sqrt{2}} (|u\bar{u}\rangle + |d\bar{d}\rangle)
  = \frac{1}{\sqrt{3}} |\eta_8\rangle + \sqrt{\frac{2}{3}} |\eta_0\rangle~,
\label{etaNSdef}
        \\
        |\eta_\St\rangle
        &=&
            |s\bar{s}\rangle
  = - \sqrt{\frac{2}{3}} |\eta_8\rangle + \frac{1}{\sqrt{3}} |\eta_0\rangle~.
\label{etaSdef}
        \end{eqnarray}

Although mathematically equivalent to the ({\it effective}) octet-singlet 
$\eta_8$--$\eta_0$ basis
(with its state mixing angle $\theta = \phi - \arctan \sqrt{2}$), 
the {\it NS--S} mixing basis and its angle $\phi$ have the advantage that 
they offer the quickest way to show the consistency of our procedures and 
the corresponding results obtained using just one state mixing angle (be 
it $\theta$ or $\phi$), with the {\bf two-mixing-angle scheme}.
This scheme (reviewed in, {\it e.g.}, Ref. \cite{Feldmann99IJMPA}) 
is defined with respect to the mixing of the four decay constants
$f^8_\eta$, $f^8_{\eta^\prime}$, $f^0_\eta$, and $f^0_{\eta^\prime}$,
and results are parameterized in terms of two auxiliary decay constants $f_0$, $f_8$, 
and two angles $\theta_0$, $\theta_8$.
However, phenomenology seems to justify the central
assumption of Feldmann, Kroll and Stech (FKS) \cite{FeldmannKrollStech98PRD}
that in the {\it NS--S} basis, the decay constants follow (in a good approximation)
the pattern of particle state mixing. As a consequence, 
$\theta_8$, $\theta_0$, $f_8$ and $f_0$ can be expressed 
\cite{FeldmannKrollStech98PRD,FeldmannKrollStech99PLB,Feldmann99IJMPA}
through the single {\it NS--S} state mixing angle $\phi$ and 
$f_\NSt$, $f_\St$, the respective decay constants of $\eta_\NSt,\eta_\St$: 
\begin{eqnarray}
\theta_8 &=& \phi - \mbox{\rm arctan}\left(\sqrt{2} f_\St/f_\NSt \right)~,
            \qquad
\theta_0 = \phi - \mbox{\rm arctan}\left(\sqrt{2} f_\NSt / f_\St \right)~,
\label{th_8th_0}
\\
f_8 &=& \sqrt{\frac{1}{3} f_\NSt^2 + \frac{2}{3} f_\St^2 }~, 
  \qquad
f_0 = \sqrt{\frac{2}{3} f_\NSt^2 + \frac{1}{3} f_\St^2 }~,
\label{f_8f_0}
\end{eqnarray}

The relations (\ref{th_8th_0})-(\ref{f_8f_0}) are applicable also in our SD-BS 
approach, as shown in detail in Appendix of Ref. \cite{Kekez:2000aw}. This reference 
confirmed our earlier results \cite{Klabucar:1997zi} on the mixing angle, 
finding that the preferred {\it NS--S} state mixing angle in our 
SD-BS approach is $\phi\approx 42\deg$, in good agreement with FKS results quoted 
in Refs. \cite{FeldmannKrollStech98PRD,FeldmannKrollStech99PLB,Feldmann99IJMPA}.
The values we find \cite{Kekez:2000aw} for $f_0$, $f_8$, $\theta_0$ and $\theta_8$ 
are also similar to theirs. This is the consequence of our values of $f_\NSt$ and 
$f_\St$, which are calculated directly from $q\bar q$ substructure, {\it i.e.}, our 
SD and BS solutions.
The ratio $y=f_\NSt/f_\St$ is suitable for giving the extent of the SU(3) breaking.
Due to Goldberger--Treiman relation for constituent quarks, it is close to 
${\cal M}_u/{\cal M}_s $, the ratio of the masses of the nonstrange and 
strange constituent quarks.  
 
We get $f_\NSt \equiv f_\pi$ ({\it theoretical} FKS analysis 
\cite{FeldmannKrollStech98PRD,Feldmann99IJMPA} assumes this),
while our numerical calculation yields $f_S = 1.451 f_\pi$,
just 3\% above the theoretical FKS prediction
\cite{FeldmannKrollStech98PRD,FeldmannKrollStech99PLB,Feldmann99IJMPA}.
This leads to $f_8 = 1.318 f_\pi$ and $f_0 = 1.170 f_\pi$.
(Interestingly, this is practically equal to our SD-BS model values
${f}_{\eta_8}=1.31 f_\pi$ and ${f}_{\eta_0}=1.16 f_\pi$ \cite{Klabucar:1997zi}
for the octet and singlet axial-current decay
constants $f_{\eta_8}$ and $f_{\eta_0}$
defined in the standard way through the matrix elements
$\langle 0 | A^{a\mu} |\eta_a\rangle$, $(a=8,0)$, so that
$f_{\eta_8}$ and $f_{\eta_0}$ are straightforwardly expressed
through $f_\NSt$ and $f_\St$ by
the relations $f_{\eta_8}= \frac{1}{3} f_\NSt + \frac{2}{3} f_\St$
and $f_{\eta_0}= \frac{2}{3} f_\NSt + \frac{1}{3} f_\St$.
We thus note that
the quadratic relations (\ref{f_8f_0}) for differently
defined octet and singlet constants $f_8$ and $f_0$, lead to similar values
as the linear relations for $f_{\eta_8}$ and $f_{\eta_0}$.)

Using in Eqs.~(\ref{th_8th_0})-(\ref{f_8f_0})
our model predictions $\phi=42\deg$ and $y=f_\NSt/f_S=0.689$,
we get the following two decay-constant-mixing angles: 
$\theta_8 = -22^\circ$ and $\theta_0 = -2.3^\circ$, close to the theoretical
FKS results \cite{FeldmannKrollStech98PRD,FeldmannKrollStech99PLB}.
See also Table 1 in Ref. \cite{Feldmann99IJMPA}, line ``FKS scheme \& theory",
giving $\theta_8 = -21.0^\circ$ and $\theta_0 = -2.7^\circ$,
while the line ``FKS scheme \& phenomenology" in this table has only
somewhat more negative $\theta_0$ but larger $f_0/f_\pi$.
Up to corrections of order $(1~-~y)^2$, the ``FKS scheme \& theory" 
implies the {\it effective} $\eta_8$-$\eta_0$ state-mixing angle 
$\theta \approx (\theta_8 + \theta_0)/2 \approx -12^\circ$, in agreement
with our results \cite{Kekez:2000aw}. Notably, $\theta \approx -12^\circ$
agrees well with the mixing angle we get from our $\eta$--$\eta'$ mass 
matrix, which agrees well with the {\it NS-S} mass matrices 
following from phenomenology and lattice calculations, as commented
in the pedagogical exposition \cite{BledWorkshop2002}. 
Reference \cite{BledWorkshop2002} also explains how our approach
practically ensures the reproduction of the empirical {\it NS-S} mass 
matrix once SD-BS calculations successfully reproduce pion and kaon 
masses and pion decay constant, as only one more parameter is then 
needed to fit $\eta$ and $\eta'$ masses satisfactorily.

\vskip 1mm
%
%
\noindent {\bf Acknowledgment:}
D. Klabu\v car thanks the organizers, B. Golli, R. Krivec, M. Rosina and 
S. \v Sirca, for their hospitality and partial support enabling his 
participation at XVIII European Conference on Few-Body Problems in Physics.
%


\end{document}